\documentclass[aps,preprint]{revtex4}
\usepackage{amsfonts}
\usepackage{amsmath}

\DeclareMathOperator{\arcsech}{sech^{-1}}

\usepackage{amssymb}
\usepackage{graphicx}
\usepackage{caption}
\usepackage{subcaption}
\providecommand{\U}[1]{\protect\rule{.1in}{.1in}}

\begin{document}
\title{Extremal Black Holes in Cosmology from Colliding Light Beams}
\author{M. Halilsoy}
\email{mustafa.halilsoy@emu.edu.tr}
\author{V. Memari}
\email{vahideh.memari@emu.edu.tr}
\affiliation{Department of Physics, Faculty of Arts and Sciences, Eastern Mediterranean
University, Famagusta, North Cyprus via Mersin 10, T{\"u}rkiye}
\date{\today}
\begin{abstract}
    From the duality between the interaction region of colliding waves and black holes (BHs) it is known that certain BH solutions can be obtained from colliding gravity coupled electromagnetic (em) waves. In the limit of vanishing gravity waves, we show that extremal Reissner-Nordstrom  (RN) BH can form. We show also that direct collision of pure em waves creates the near horizon geometry (NHG) of the same BH, supporting our point. Due to the quantum restrictions such BHs can have horizon radii in the range $R>10^{8}m$. If such BHs from pure light do exist this may open a new frontier in our understanding of cosmological BHs at large. The generation of impulsive gravitational waves in the process may be instrumental in the detection of a BH created from pure light.
\end{abstract}
\maketitle
 \section{Introduction}
 Colliding waves was a well-established, subject of classical general relativity (GR) during the 70s and 80s \cite{1,2,3,4}. While the collision of pure gravitational waves was doomed by a spacelike singularity, the collision of pure em waves proved regular apart from the emergent null-singularities on the light cone. Chandrasekhar and Xanthopoulos (CX) \cite{5} had established a duality between colliding wave spacetimes and BHs. The main cause for revisiting this old topic is to check whether, from the collision of pure em waves, possible BH emerges or not. From the outset, we state that the original solution of colliding em shock waves due to Bell-Szekeres (BS) \cite{1} is not a BH solution, since its horizon is not an event horizon \cite{6}. For this reason, we approach the problem from two different methods: we consider first colliding Einstein-Maxwell (EM) waves which contain both em and gravitational waves by setting at the end the gravity waves to zero. As our second method we show that colliding pure em waves spacetime coincides with the NHG \cite{7,8} of the extremal RN BH. It has been proved recently that the formation of BHs from pure light with horizon radii in the interval $10^{-29}m<R<10^{8}m$ is not possible \cite{9}. The argument is based on the formalism established by Schwinger in quantum electrodynamics (QED) \cite{10}. As described in great detail the particle creation causing loss of energy prevents the huge accumulation of mass/energy and in consequence, a BH does not form \cite{9}. That is, as the quantum revolution at the beginning of the 20th century saved electrons from collapsing onto the nucleus of the H-atom, similar principles save pure light also from collapsing into BH. Since the argument is valid only for pure em fields inclusion of gravity as initial input must be excluded. No doubt, the gravity of any form serves towards collapse either into BHs or naked singularities. For this reason, we consider first, in section II the collision problem of gravity plus em waves to see the formation of a BH. Next, as we are interested in pure em waves we recall the geometry of colliding em waves. In this case, prior to the collision, we have pure em waves whereas, in the post-collision region i.e. the interaction region, we have emergent impulsive gravitational waves on the null boundaries. Off the null cone, the spacetime of the interaction region is conformally flat, transformable to the Bertotti-Robinson (BR) spacetime \cite{11,12}. It is known also that BR spacetime is the NHG of the extremal RN BH. This particular point provides our second argument after collision of gravitation $+$ em wave problem towards the formation of a BH from pure light. Given the Schwinger limits of $10^{-29}m<R<10^{8}m$, we show that the lower bound gives a BH with physically non-acceptable parameters. For instance, the charge density of a possible BH turns out to be of the order $\sim 10^{76}C/m^{3}$, whereas, for an electron, this figure is $\sim 10^{24}C/m^{3}$. Formation of such a BH from physical grounds is not acceptable. The upper Schwinger bound, $R>10^{8}m$, however, will be shown to admit an acceptable BH, as a result of two colliding em waves at a large scale. Namely, two intersecting highly energetic $\gamma-$ ray bursts can give rise to the formation of such a BH. We recall that the upper bound energy for $\gamma-$ rays is not limited, and we make use of this as an advantage. In brief, the huge em radiation emitted $\sim380000$ years after the Big Bang and their mutual collision may create the primordial BHs from pure light in our universe to explain also early galaxy formation centered by early BHs. However, according to our calculations, there is no room for micro BHs from pure light. Our paper is organized as follows: In section II, we consider BH formation via isometric transformation from colliding EM waves. In section III, we concentrate on the pure em collision problem, and in section IV we complete our paper with a conclusion and discussion. 
 \section{Black Holes from Colliding Einstein-Maxwell waves}
 We start with the line element
 \begin{equation}
     ds^{2}=\left(1+\alpha\sin\left(au+bv\right)\right)^{2}\left(2dudv-\cos^{2}\left(au-bv\right)dx^{2}\right)-\frac{\cos^{2}\left(au+bv\right)}{\left(1+\alpha\sin\left(au+bv\right)\right)^{2}}dy^{2}\label{1}
 \end{equation}
 where $\left(a,b\right)$ are energy constants, $\alpha$ is a parameter with $0\leq\alpha\leq1$, and the null coordinates $\left(u,v\right)$ are to be understood with the Heaviside step function, i.e. $u\rightarrow u\Theta\left(u\right)$ and $v\rightarrow v\Theta\left(v\right)$. Such substitutions are essential for studying the wave collision problem in GR \cite{2}.  This spacetime is the linearly polarized version of the more general case of CX \cite{5}. The fact that this line element represents the collision of EM waves is explained in great detail in \cite{5}. In order to realize the physical content of this spacetime we make use of the Newman-Penrose (NP) \cite{13} null-tetrad formalism. An appropriate tetrad is provided by the set of the following null vectors
 \begin{align}
     l_{\mu}&=\left(1+\alpha\sin\left(au\Theta\left(u\right)+bv\Theta\left(v\right)\right)\right)\delta^{u}_{\mu}\nonumber\\
     n_{\mu}&=\left(1+\alpha\sin\left(au\Theta\left(u\right)+bv\Theta\left(v\right)\right)\right)\delta^{v}_{\mu}\label{2}\\
     \sqrt{2}m_{\mu}&=\left(1+\alpha\sin\left(au\Theta\left(u\right)+bv\Theta\left(v\right)\right)\right)\cos\left(au\Theta\left(u\right)-bv\Theta\left(v\right)\right)\delta^{x}_{\mu}+\frac{i\cos\left(au\Theta\left(u\right)+bv\Theta\left(v\right)\right)}{1+\alpha\sin\left(au\Theta\left(u\right)+bv\Theta\left(v\right)\right)}\delta^{y}_{\mu}\nonumber\\
     \bar{m}_{\mu}&=\left(\textit{complex conjugate of }m_{\mu}\right)\nonumber
 \end{align}
 The energy-momentum tensor $T_{\mu\nu}$, which satidfies the Einstein equations $G_{\mu\nu}=R_{\mu\nu}-\frac{1}{2}Rg_{\mu\nu}=-T_{\mu\nu}$, in the present problem is given by
 \begin{equation}
     \frac{1}{2}T_{\mu\nu}=\Phi_{22}l_{\mu}l_{\nu}+\Phi_{00}n_{\mu}n_{\nu}+\Phi_{20}m_{\mu}m_{\nu}+\Phi_{02}\bar{m}_{\mu}\bar{m}_{\nu}\label{3}
 \end{equation}
 with all other components vanishing due to the assumed symmetries. Together with the non-zero tetrad components of the Weyl curvatures $\Psi_{A}$, we have the following table of NP quantities:
 \begin{align}
    \Phi_{00}&=\left(1-\alpha^{2}\right)\frac{b^{2}\Theta\left(v\right)}{\Sigma}\nonumber\\
    \Phi_{22}&=\left(1-\alpha^{2}\right)\frac{a^{2}\Theta\left(u\right)}{\Sigma}\nonumber\\
    \Phi_{02}&=\Phi_{20}=\left(1-\alpha^{2}\right)\frac{ab\Theta\left(u\right)\Theta\left(v\right)}{\Sigma}\label{4}\\
    \Psi_{2}&=\frac{\alpha\left(\alpha+\sin\left(au+bv\right)\right)}{\Sigma}ab\Theta\left(u\right)\Theta\left(v\right)\nonumber\\
    \Psi_{0}&=\delta\left(v\right)F_{1}\left(u\right)+\frac{3\alpha b^{2}\Theta\left(v\right)}{\Sigma}\left[\alpha+\sin\left(au\Theta\left(u\right)+bv\Theta\left(v\right)\right)\right]\nonumber\\
    \Psi_{4}&=\delta\left(u\right)F_{2}\left(v\right)+\frac{3\alpha a^{2}\Theta\left(u\right)}{\Sigma}\left[\alpha+\sin\left(au\Theta\left(u\right)+bv\Theta\left(v\right)\right)\right]\nonumber
 \end{align}
 where $\Sigma=\left(1+\alpha\sin\left(au+bv\right)\right)^{4}$, $\delta\left(u\right)$, $\delta\left(v\right)$ are Dirac delta functions and $F_{1}\left(u\right)$, $F_{2}\left(v\right)$ are well-defined functions that they vanish for $\alpha=0$, prior to the collision of em waves. Namely, for $\alpha=0$ we have
 \begin{equation}
     F_{1}\left(u\right)=-\frac{b\tan\left(au\Theta\left(u\right)\right)}{\left(1+\sin\left(au\Theta\left(u\right)\right)\right)^{3}}\label{5}
 \end{equation}
 which vanishes prior to the collision and a similar argument is valid also for $F_{2}\left(v\right)$. For $u>0$ and $v>0$, however, we have nonzero impulsive gravitational waves even for the pure em collision problem, which is well-known for a long time \cite{1}. All our arguments are valid for the 'inside' region of the interaction region, that is for $u>0$ and $v>0$. We note also that in the interaction region, off the null boundaries, we have the type-D condition $9\Psi_{2}^{2}=\Psi_{0}\Psi_{4}$, satisfied. Now, we proceed by identifying the BH in the interaction region. For this reason, we introduce new coordinates $\left(\eta,\mu\right)$  defined as follows
\begin{align}
    \eta&=\sin\left(au+bv\right)\nonumber\\
    \mu&=\sin\left(au-bv\right)\label{6}
\end{align}
so that the following relation
\begin{equation}
    \frac{d\eta^{2}}{1-\eta^{2}}-\frac{d\mu^{2}}{1-\mu^{2}}=4ab dudv\label{7}
\end{equation}
holds. For $\alpha=1$ it is not difficult to show that the transformation
\begin{align}
    \eta&=\frac{r}{m}-1\nonumber\\
    \mu&=\cos{\theta}\nonumber\\
    x&=\frac{1}{\sqrt{2ab}}\varphi\label{8}\\
    y&=\frac{t}{m\sqrt{2ab}}\nonumber
\end{align}
with appropriate scalings on $\left(x,y\right)$ gives us the Schwarzschild metric with mass $m$ as the scaling constant in the form
 \begin{equation}
     \left(2ab\right)m^{2}ds^{2}=\left(1-\frac{2m}{r}\right)dt^{2}-\frac{dr^{2}}{1-\frac{2m}{r}}-r^{2}\left(d\theta^{2}+\sin^{2}\theta d\varphi^{2}\right)\label{9}
     \end{equation}
     In other words, $\alpha=1$ corresponds to the problem of colliding pure gravitational waves, where it was shown \cite{3} that it consists of impulsive and shock gravitational waves. Consideration of $0<\alpha<1$ gives us the collision of coupled gravitational and em waves \cite{5}. Herein our main interest is the case of the formation of a BH from pure em waves. To this end, we revise the transformation of (\ref{8}) as follows
\begin{align}
    \eta&=\frac{1}{\alpha}\left(\frac{r}{m}-1\right)\nonumber\\
    \mu&=\cos{\theta}\nonumber\\
    x&=\frac{1}{\sqrt{2ab}}\varphi\label{10}\\
    y&=\frac{\alpha t}{m\sqrt{2ab}}\nonumber
\end{align}
It can easily be checked that through (\ref{10}) and with necessary scaling of the $x,y$ coordinates one obtains 
\begin{equation}
     \left(2ab\right)m^{2}ds^{2}=Fdt^{2}-\frac{dr^{2}}{F}-r^{2}\left(d\theta^{2}+\sin^{2}\theta d\varphi^{2}\right)\label{11}
     \end{equation}
where
\begin{equation}
    F=1-\frac{2m}{r}+\frac{\left(1-\alpha^{2}\right)m^2}{r^2}\nonumber
\end{equation}
This is precisely the RN form of the metric with the definition of charge (see also \cite{5} for this argument) given by
\begin{equation}
    Q^{2}=\left(1-\alpha^{2}\right)m^{2}\label{12}
\end{equation}
Now, the critical limit $\alpha\rightarrow0$, can be applied consistently to give
\begin{equation}
    F=\left(1-\frac{m}{r}\right)^{2}\label{13}
\end{equation}
in (\ref{11}) to give the metric of an extremal RN BH. In the next section, we show that the direct transition from the colliding em wave metric to the Schwarzschild coordinates gives the NHG of the extremal RN BH, which is also another evidence for a BH from colliding pure em waves.  
 \section{Electromagnetic BHs}
 It is known from GR, with textbook clarity \cite{1,2,3}, that interaction (collision) of pure em waves creates a spacetime that through NHG becomes equivalent to the geometry of extremal RN BH \cite{7,8}. To show this we recall the line element of extremal RN BH
 \begin{equation}
     ds^{2}=\left(1-\frac{|Q_{g}|}{r}\right)^{2}dt^{2}-\left(1-\frac{|Q_{g}|}{r}\right)^{-2}dr^{2}-r^{2}d\Omega^{2}\label{14}
 \end{equation}
where in the geometrical units $M_{g}\textit{(mass)}=|Q_{g}|\textit{(charge)}$ and $d\Omega^{2}=d\theta^{2}+\sin^{2}{\theta}d\varphi^{2}$. The NHG, i.e. the smoking gun signal for the extremal RN BH, is obtained by using first the transformation $r=M+\rho$ with $\rho<<1$. Letting next $\rho=\frac{1}{R}$ leads us (after the scaling of time $T=M^{-2}t$) to the conformally flat BR metric
\begin{equation}
    ds^{2}=Q_{g}^{2}\left(\frac{dT^{2}-dR^{2}}{R^{2}}-d\Omega^{2}\right)\label{15}
\end{equation}
This is the transform of the spacetime of two colliding em waves \cite{1,2} described by the BS line element \cite{1} (take $\alpha=0$ in (\ref{1}))
\begin{equation}
    ds^{2}=2dudv-\cos^{2}{\left(au\Theta\left(u\right)-bv\Theta\left(v\right)\right)}dx^{2}-\cos^{2}{\left(au\Theta\left(u\right)+bv\Theta\left(v\right)\right)}dy^{2}\label{16}
\end{equation}
where $\left(a,b\right)$ are energy constants of the waves and $\Theta\left(u\right)$, $\Theta\left(v\right)$ are the step functions. With reference to \cite{2} we give the explicit transformation as follows
\begin{align}
    T+R&=\coth{\left(\frac{1}{2}\arcsech\cos\left(au+bv\right)-\frac{y}{2Q}\right)}\nonumber\\
     T-R&=-\tanh{\left(\frac{1}{2}\arcsech\cos\left(au+bv\right)+\frac{y}{2Q}\right)}\label{17}\\
     \theta&=au-bv+\frac{\pi}{2}\nonumber\\
     \varphi&=\frac{x}{2}\nonumber
\end{align}
We recall once more that radiation fields in GR can best be described by the null-tetrad formalism of NP \cite{13}. In the tetrad we have chosen in section II (with $\alpha=0$) we have the only nonzero Ricci $\left(\Phi_{ij}\right)$ and Weyl $\left(\Psi_{A}\right)$ components
\begin{align}
    \Phi_{22}&=a^{2}\Theta\left(u\right)\nonumber\\
    \Phi_{00}&=b^{2}\Theta\left(v\right)\nonumber\\
    \Phi_{02}&=ab\Theta\left(u\right)\Theta\left(v\right)\label{18}\\
    \Psi_{4}&= a\tan\left(bv\right)\Theta\left(v\right)\delta\left(u\right)\nonumber \\
    \Psi_{0}&=-b\tan\left(au\right)\Theta\left(u\right)\delta\left(v\right)\nonumber
\end{align}
The line element (\ref{15}) transforms into (\ref{16}) \cite{2} through the transformation (\ref{17}), provided $Q_{g}^{2}=\left(2ab\right)^{-1}$. Combined with the extremality condition in the geometrical units these are related as follows
\begin{equation}
    R=M_{g}=|Q_{g}|=\left(2ab\right)^{-\frac{1}{2}}\label{19}
\end{equation}
From the line element (\ref{16}) we impose the conditions $au\leq 1$ and $bv\leq 1$, and for technical simplicity, we choose symmetric collision so that $a=b$. The conditions (\ref{19}) imply that given the horizon radius $R$ we can determine all the remaining parameters. The next step is to transform the physical parameters into SI unit system in accordance with
\begin{align}
    M&=\frac{c^{2}}{G}M_{g}\nonumber\\
    Q&=c^{2}Q_{g}\left(\frac{4\pi\epsilon_{0}}{G}\right)^{\frac{1}{2}}\label{20}
\end{align}
The minimum energy required to form such a BH is $E_{\textit{min}}=Mc^{2}$. Using first the lower Schwinger bound $R=10^{-29}m$ with a charge density $\frac{3Q}{4\pi R^{3}}\sim 10^{76}C/m^{3}$ suffices to discard such a BH as impossible. We consider next the case $R=10^{9}m$ which is above the Schwinger's upper bound. This gives us
\begin{align}
    M&\sim 10^{36}kg\nonumber\\
    Q&\sim 10^{26}C \label{21}\\
    E_{\textit{min}}&\sim 10^{53}J\nonumber
\end{align}
From the flux $F=\frac{a^{2}c^{5}}{8\pi G}$, the initial em flux of the colliding waves is $F\sim10^{33}W/m^{2}$. These figures are not unusual at a cosmological scale. 
\section{Conclusion and Discussion}
In this analysis we do not present any new solution to Einstein's equations, we just make use of the previously well-known spacetime of two colliding em shock or gravity coupled em waves; so that the mutual focusing of waves creates a strong curvature and ultimate BHs. Vanishing of the gravity waves results in a BH from pure light. It is also shown that highly energetic em waves collide to form a spacetime that is isometric to the NHG of an extremal, charged BH. This amounts to an indirect evidence of a BH based upon the existence of a NHG. Due to restrictions from QED, \cite{9} towards the formation of a Kugelblitz, i.e. ball-lightning in German, the event horizon of a possible BH must satisfy $R>10^{8}m$. We have worked out explicitly the case of $R=10^{9}m$,  where the mass and charge are respectively, $M\cong 10^{36}kg\left(\approx5\times10^{5}M_\odot\right)$ and $Q\cong10^{26}C$. The corresponding $\gamma-$ rays that take part in the collision for this formation have fluxes of the order $F=10^{33}W/m^{2}$. We note that the superposition principle also serves to build up such high fluxes \cite{14}. Besides light collision it is known that primordial BHs can be produced from the collision of particles \cite{15}. Formation of a BH from a pure em wave was considered also in \cite{16}.
\par
A natural question is the following: how can we distinguish an em BH from BHs formed otherwise? Our analysis suggests that the emergent impulsive gravitational waves,i.e. $\Psi_{0}$ and $\Psi_{4}$, may serve as precursors of such exceptional, extremal BHs. We finally add that upon publication of \cite{9}, namely the absence of BHs from pure light, papers appeared \cite{17,18} with responses \cite{19} showing that this is going to be a rather hot topic. Further considerations of the BS spacetime and whether a rotating pure em BH exists from colliding cross-polarized waves \cite{20,21} are the projects to be handled in the future. The rotation factor in an em universe is known to create conformal curvature which admits interesting geodesics \cite{22}.
\section*{Data Availability Statement}
Data sharing is not applicable to this article as no new data were created or analyzed in this study.

\end{document}